\documentstyle[11pt]{article} 
\input epsf

\newcommand{\be}{\begin{equation}}
\newcommand{\ee}{\end{equation}}

\newcommand{\cluster}{{\cal C}}

\newcommand{\setg}{{\cal G}}
\newcommand{\calz}{{\cal Z}}
\newcommand{\Oc}{{\cal O}}
\newcommand{\znum}{{\sf Z \!\!\! Z}}

\newcommand{\bea}{\begin{eqnarray}} 
\newcommand{\eea}{\end{eqnarray}}
 
\begin{document}

\begin{center}
{\LARGE{\bf A new approach to significantly reduce the negative sign problem in quantum Monte Carlo.}}\\[1cm]

\vspace*{1cm}
{\Large A. Galli\footnote{galli@mppmu.mpg.de}}\\[0.3cm]
{\small \em Max-Planck-Institut f\"ur Physik, D-80805 Munich, Germany}\\[1cm]

{\bf Abstract}\\[0.3cm]
\end{center}{\small
We present a new approach for Monte Carlo simulations of lattice quantum spin systems which is able to eliminate the negative sign problem. Its complexity is linear in the volume of the lattice.  
Its efficiency is tested on a simple 2-dimensional fermionic model. 
\\[1cm]
}

The investigation of quantum spin systems is important for understanding the physics related to strongly correlated electrons and high temperature superconductivity. A poverful technique to numerically study quantum spin systems is the quantum Monte Carlo
 (QMC) method (for a review see \cite{dagotto}) based on the Suzuki-Trotter formula \cite{suzuki}. Unfortunately, QMC can suffer from the negative sign problem (NSP), which becomes exponentially serious on large lattices and at low temperature.     
In this work we present  a new algorithm which is able to strongly reduce this problem. A more detailed description of this algorithm and a rigorous proof of its correctness will be published elsewhere
\cite{QM}. This algorithm is general and can be applied to any
quantum spin system. Its complexity is only linear in the volume of the lattice. We have tested its efficiency with a simple 2-dimensional fermionic model and we show that for this model the NSP disappears.\\

 We consider a quantum spin system defined on a d-dimensional finite lattice.
 Its direct simulation is not possible because the requirement in storage and work is exponential in the lattice size of the system. To avoid this complexity, usually one transforms it to a d+1-dimensional classical spin system realized by the application
 of the Suzuki-Trotter formula to the original d-dimensional system. The extra dimension is a discretization of the inverse temperature. We call it time. This classical spin system allows us to use a Monte Carlo approach for evaluating observables. The co
nfigurations $v$ defined on the d+1-dimensional lattice are weighted by $w(v)$ in the partition function
 \be
 Z=\sum_v w(v)
 \ee
 The evaluation of the weight $w(v)$ is of small complexity, contrary to the original d-dimensional quantum spin system.
 The price to pay is that the new classical system has a partition function for which generally not all weights are positive semidefinite. This fundamental difficulty is usually referred to as the "negative sign" problem. It is not related to any approxim
ations in the Monte Carlo scheme but it describes the fact that the statistical error of the observables can become very large, increasing exponentially in the inverse temperature $\beta$ and lattice volume.\\

 Any classical observable A is measured at the first time slice of the d+1-dimensional lattice by averaging over the sample of configurations generated by the Monte Carlo process. For simplicity, we restrict the discussion to classical observables. The ex
pectation value of them can be written
 \be
 <A>_w= \frac{\sum_v A(v)\,|w(v)|\cdot sgn(w(v))}{\sum_v |w(v)|}\times \frac{1}{<sgn(w(v))>_{|w|}}
 =:\frac{<A\,sgn>_{|w|}}{<sgn>_{|w|}}\label{evplus}
 \ee
 where $<...>_w$ and $<...>_{|w|}$ denote the average taken with the weights $w$ and $|w(v)|$, respectively. If the average sign is small there will be large cancelations making an accurate evaluation of $<A>$ impossible so that meaningful results are dif
ficult to obtain.\\

 We denote a Monte Carlo algorithm $\phi_{|w|}$ which generates configurations $v$ in equilibrium with the distribution $|w(v)|$. 
 This algorithm can be realized in different ways. The explicit realization of it is for the moment irrelevant to the discussion. We denote the expectation value measured with the algorithm $\phi_{|w|}$ by eq. (\ref{evplus}).
 It is important to notice that the value $A(v)$ of an observable $A$ does not depend on the complete configuration $v$ but only on the part living on the time slice at $t=0$.  This is a crucial property that we will use for constructing an algorithm whic
h is able to reduce the sign problem by substantially increasing the average sign. \\

 We consider a mapping $g$ which maps a configuration $v$ to a configuration $gv$ and which satisfies the following condition: for any observable $A$ we have $A(v)=A(gv)$.  Such a mapping is easily constructed:  we can require that it does not change the 
configuration $v$ at $t=0$. We consider a set $\setg$ of such mappings with the additional property that these mappings are bijective.  
 It is easy to prove \cite{QM} that using the weight defined by
 \be
 \tilde w(v,g)=\frac{1}{2}(w(v)+w(gv))\label{tildew}
 \ee
 we obtain for the expectation value (\ref{evplus}) of an observable 
 \be
 \frac{\sum_v A(v)w(v)}{\sum_v w(v)}=
 \frac{\sum_{g\in\setg}\sum_v A(v)\tilde w(v,g)}{\sum_{g\in\setg}\sum_v \tilde w(v,g)}
 \label{newev}
 \ee
 Using this new weight for a simulation it is clear that the expectation value of $A$ measured with a Monte Carlo algorithm $\phi_{|w|}$ with state space $\{v\}$ and equilibrium distribution $|w(v)|$ is the same as the one measured with a Monte Carlo algo
rithm $\phi_{|\tilde w|}$ with state space $\{(v,g)\}$ and equilibrium distribution $|\tilde w(v,g)|$.\\

 We suppose that we can construct these mappings $g\in \setg$ so that the average of the sign if measured with $\phi_{|\tilde w|}$ satisfies 
 \be
 <sgn(v)>_{|\tilde w|}\,\,\,\,\geq\,\,\,\,<sgn(v)>_{|w|}
 \label{suppression}
 \ee
 where $<...>_{|\tilde w|}$ and $<...>_{|w|}$ mean the expectation taken from the $|\tilde w|$ and $|w|$ distributions, respectively. 
 In the case that the average sign {\em substantially} increases then it is evident that the sign problem is suppressed.\\

  Such a mapping can be realized by flipping clusters of spins in $v$ such that if $w(v)\neq 0$ then also $w(gv)\neq 0$ and the following condition is satisfied
\be
\mbox{{\sf condition}}=\{(w(v)+w(\Xi_\cluster v)>0)\mbox{ {\bf and} $\Oc(v,\Xi_\cluster v)$}\}\label{co}
\ee
 where $\Xi_\cluster$ denotes the flipping of the spins belonging to the cluster $\cluster$. The first part of the {\sf condition} is needed to obtain a positive weight $\tilde w(v,g)$. The second part of the condition $\Oc(v,\Xi_\cluster v)$ takes the va
lue {\bf true} if the flipping $\Xi_\cluster$ maps bijectively the configuration $v$ to $gv$, otherwiese it takes the value {\bf false}. This part of the {\sf condition} is realised with very high accurancy using a hashing technique \cite{hash}. To be spe
cific, the boolean function $\Oc(v,v')$ is defined by the following procedure
 {\tt
 \bea
 \Oc(v,v'):&&
 \mbox{if $H_{table}(h(v'))=0$ then $H_{table}(h(v')):=h(v)$ endif}\nonumber\\
 &&\mbox{if $H_{table}(h(v'))\neq h(v)$ then output $\Oc$={\bf false}}\nonumber\\
 &&\mbox{else output $\Oc$={\bf true} endif}\nonumber
 \eea}
 where $h$ is a hashing function which assigns in an arbitrary way a non vanishing integer label (in an interval $[1,\calz]\cap\znum$) to any state $v$, and $H_{table}$ is a hashing table with $\calz$ integer entries.

 The search of the clusters follows a fixed order of operations. One can select a point in the d+1-dimensional lattice from a fixed list of points and construct a cluster starting from it. During the construction a fixed list of random numbers can be used
. It is important, however, that the two lists always remain the same every time one applies this procedure to a configuration. Changing the lists is equivalent to select a new mapping in $\setg$. If the flip of the constructed cluster does not satisfy th
e {\sf condition} (\ref{co}) the next point in 
 the list can be selected and a new cluster constructed and this search is repeated until the {\sf condition} (\ref{co}) is satisfied. If the {\sf condition} (\ref{co}) is never satisfied the procedure can be stopped, when all the points in 
 the list are tested once, and the original state is returned as the result.
 This algorithm defines a mapping $g\in\setg$. The complexity of the search of a cluster is in this way linear in the volume of the lattice.
 The flip of the cluster $\Xi_\cluster$ is defined so that $A(v)=A(\Xi_\cluster v)$ for any observable $A$.\\

 This mapping defines our new weight $\tilde w$. A Monte Carlo algorithm $\phi_{|\tilde w|}$ can now be used to update the configurations and the mappings with respect to the new weight. 
 The first part of the {\sf condition} (\ref{co}) guarantees us that 
 after average (\ref{suppression}) is satisfied. 
 The second part of the {\sf condition} (\ref{co}) guarantees us that
 the mapping $g$ maps a configuration $v$ to $gv$ bijectively with probability $1-O\left(\frac{1}{\calz}\right)$. Since the mapping is not always bijective, during the Monte Carlo process a systematic error is produced with probability 
$O\left(\frac{1}{\calz}\right)$. The systematic error in the average (\ref{evplus}) is then after average over the collected sample of configurations of order $\frac{1}{\calz}$. This systematic error can be tuned in the algorithm by increasing the dimensi
on $\calz$ of the hashing table.  We can chose its dimension so that $\frac{1}{\calz}$ is much smaller than the statistical error. In addition the systematic error can be measured by counting the collisions in the hashing table.\\
 
If the set of mappings $\setg$ contains only one element then to avoid too much collisions in the hashing the dimension of the hashing table has to be much larger than the number of Monte Carlo iterations one desires to perform. This is doable but require
s a lot of memory. If, however, the set of mappings $\setg$ contains a huge amount of elements then the hashing table can be small because the algorithm uses a selected mapping only for a short time and then selects a new one according to the distribution
 $|\tilde w(v,g)|$ (using a Metropolis acceptance test). If the set $\setg$ is big enough, the probability that the algorithm selects the same mapping twice is 
infinitesimal so that there is no need to store the history of the hashing tables of old mappings.\\

As an example we consider a simple model of free quantum spins in a chemical potential. This example was analyzed by quantum Monte Carlo using the loop algorithm \cite{evertz} in \cite{wiese}. 
We consider fermions living on the sites of a spatially 2-dimensional
square lattice with periodic boundary conditions. 
We consider a particle with only one internal degree of freedom.
Creation and annihilation operators
$c_x^*$ and $c_x$ anticommute. The Hamilton operator is
\begin{equation}
H = \sum_{x,i} (c_x^* c_x + c_{x+\hat{i}}^* c_{x+\hat{i}}
- c_x^* c_{x+\hat{i}} - c_{x+\hat{i}}^* c_x),
\end{equation}
where $\hat{i}$ is the unit vector in the $i$-direction. The model is trivial and
can be solved in momentum space. However, when one tries to simulate it with a Monte Carlo algorithm it shows from the algorithmic point of view many of the 
characteristic features of more complicated quantum spin systems.
In addition, the exact solution of this model allows us to test our algorithm. 
We can express the grand canonical partition function 
\be
Z=Tre^{-\beta(H-\mu N)}
\ee
(where $N$ is the particle number operator)
with the Suzuki-Trotter formula as explained in \cite{wiese,QM}.\\

In Table 1 we present the results of the Monte Carlo simulations performed with the loop algorithm and with our algorithm and we compare the obtained results with the exact solution.
The construction of the clusters in our algorithm 
can be performed in analogy with the loop algorithm \cite{wiese}, however using a fixed list of points in the d+1-dimensional lattice and a fixed list of random numbers for each mapping $g\in \setg$. For our runs we have used a hashing table with $\calz=1
0000$ entries. In all our simulations we have not found any collisions in the hashing tables so that the systematic error on the observables is zero.\\

We have applied the algorithms for various values of $\beta$ and $\mu$ at fixed lattice spacing $\beta\epsilon=1/16$. The results of both algorithms agree with the exact results within the error bars. It is evident that the sign problem becomes severe for
 the loop algorithm when the temperature is lowered or the chemical potential is increased. However, the sign remains always positive for our algorithm. \\

A more exhaustive analysis of the efficiency of our algorithm is under study. Applications of it to more physical interesting models are under way.\\

{\Large {\bf Acknowledgments}}\\

I would like to specially thank N. Galli for discussions and help with the hashing technique. 
I also would like to thank B. Jegerlehner for helpful comments and 
discussions and P.Weisz for reading the manuscript and helpful comments.

\begin{table}
\begin{center}
{\footnotesize
\begin{tabular}{|c|c|c|c|c|c|c|c|c|}\hline
$\beta$&$\mu$&$\sqrt{|\Lambda|}$&$4T$&$<n>_O$&$<n>_L$&$<n>_E$&$<sgn>_O$&$<sgn>_L$\\\hline\hline
0.5 & 2 & 8 & 32 & 0.304(2) & 0.305(3) & 0.3049 & 1.0(0) & 0.690(4) \\\hline
0.5 & 4 & 8 & 32 & 0.501(2) & 0.498(3) & 0.5000 & 1.0(0) & 0.591(4) \\\hline
1.0 & 2 & 8 & 64 & 0.233(2) & 0.230(40) & 0.2321 & 1.0(0) & 0.048(6)\\\hline
1.0 & 3 & 8 & 64 & 0.357(2) & ? & 0.3568 & 1.0(0) & -0.003(6)\\\hline
1.0 & 4 & 8 & 64 & 0.499(2) & ? & 0.5000 & 1.0(0) & 0.002(8)\\\hline
2.0 & 2 & 8 & 128 & 0.193(2) & ? & 0.1956 & 1.0(0) & 0.004(9)\\\hline
2.0 & 4 & 8 & 128 & 0.498(2) & ? & 0.5000 & 1.0(0) & -0.001(9)\\\hline\hline
1.0 & 2 & 12 & 64 & 0.232(2) & 0.20(10) &  0.2321 & 1.0(0) & 0.018(6)\\\hline
1.0 & 3 & 12 & 64 & 0.358(2) & ? & 0.3568 & 1.0(0) & 0.001(3) \\\hline
1.0 & 4 & 12 & 64 & 0.501(2) & ? & 0.5000 & 1.0(0) & -0.001(4)\\\hline\hline
\end{tabular}}
\end{center}
\caption{{\small Results from the Monte Carlo simulations 
for different parameters. The simulations are done with
the loop algorithm $\phi_{|w|}$ and our algorithm. 
The expectation values of the number operator $n$ and the sign obtained with the loop algorithm are denoted by $<n>_L$ and $<sgn>_L$, the ones 
obtained with our algorithm are denoted by $<n>_{O}$ and $<sgn>_O$. The mark "?" indicates that the measurement of the expectation value was impossible, because the statistical error exceeds the expectation value. The exact values of $<n>$ are denoted by 
$<n>_E$. 
}}
\end{table}


\begin{thebibliography}{99}
\bibitem{dagotto} E.Dagotto, Rev. Mod. Phys. 66 (1994) 763\\
W.van der Linde, Phys. Rep. 220 (1992) 53
\bibitem{suzuki}M.Suzuki, Prog. Theor. Phys. 56 (1976) 1454; J. Stat. Phys 43 (1986) 883
\bibitem{QM} A.Galli, in preparation.
\bibitem{hash} For an introduction see: D.E.Knuth, The art of computer programming, Addison-Wesley, (1973), Vol. 3, pp. 506-550
\bibitem{evertz}H.G. Evertz, Nucl. Phys. B (Proc. Suppl.) 30 (1993) 277\\
H.G. Evertz, G. Lana and M. Marcu, Phys. Rev. Lett. 70, 875 (1993)
\bibitem{wiese}U.J. Wiese, Phys. Lett. B311 (1993) 235
\end{thebibliography}
\end{document}